\documentclass[fleqn,usenatbib]{mnras}

\usepackage{newtxtext,newtxmath}

\usepackage[T1]{fontenc}

\DeclareRobustCommand{\VAN}[3]{#2}
\let\VANthebibliography\thebibliography
\def\thebibliography{\DeclareRobustCommand{\VAN}[3]{##3}\VANthebibliography}

\usepackage{graphicx}	
\usepackage{amsmath}	
\usepackage{amssymb}	

\title[X-ray Observations of Two Type Ia SNe]{X-ray Observations of Two Type Ia Supernovae with an  H$\alpha$ Line in their Optical Spectrum}

\author[V. V. Dwarkadas]{
Vikram V. Dwarkadas$^{1}$\thanks{E-mail: vikram@astro.uchicago.edu}
\\
$^{1}$Department of Astronomy and Astrophysics, University of Chicago, 5640 S Ellis Ave., ERC 569, Chicago, IL 60637
}

\date{Accepted XXX. Received YYY; in original form ZZZ}

\pubyear{2022}

\begin{document}
\label{firstpage}
\pagerange{\pageref{firstpage}--\pageref{lastpage}}
\maketitle

\begin{abstract}
We report on Chandra X-ray observations of  SN 2016jae and SN 2018cqj, both low luminosity Type Ia supernova that showed the presence of a H line in their early optical spectrum. No X-ray emission is detected at the location of either SN. Upper limits to the luminosity of up to 2 $\times 10^{40}$\,erg s$^{-1}$ are calculated for each SN, depending on the assumed spectral model, temperature and column density. This luminosity is comparable to that of another low-luminosity Type Ia SN, SN 2018fhw, that was observed with Chandra. It is generally lower than upper limits calculated for Type Ia-CSM SNe observed in X-rays, and also below that of SN 2012ca, the only Type Ia-CSM SN to have been detected in X-rays. Comparisons are made to other Type Ia SN with a H line observed in X-rays. The observations suggest that while the density into which the SN is expanding may have been high at the time the H$\alpha$ line was detected, it had decreased considerably by the time of X-ray observations. 
\end{abstract}




\section{Introduction}

The classification of supernovae (SNe) into two types, Type Ia and Type II, mainly happens on the basis of their optical spectra, with sub-classifications also based on their light curves. At the most basic level, those SNe that show a H line in their optical spectrum are known as those of Type II, and those that do not show H fall into the Type I category. The latter are further divided in Ia, Ib and Ic, with the Ia's showing a distinct Si line in their spectra, which is missing in the Ib/cs. Although the Type Ia's constitute a small fraction of all SNe, they have gained importance as cosmological `candles' used to measure the expansion and acceleration of the universe. 

Given their importance, it is at the same time concerning that the progenitors of Type Ia SNe are not well known \citep{lrh23}. The progenitor is accepted to be a white dwarf. If a white dwarf gains mass from a binary companion and exceeds the Chandrasekhar limit, it can explode as a Type Ia. Alternately, two white dwarfs may merge together producing a Type Ia. It is the nature of the companion that therefore inspires considerable debate \citep{pilar14}. Evidence exists for both double-degenerate and single-degenerate systems.  In the double-degenerate scenario, the companion is another white dwarf \citep{scalzoetal10, silvermanetal11, nugentetal11, bloometal12, brownetal12}, while in the single-degenerate case it is a main sequence or evolved star \citep{hamuyetal03, dengetal04}. Difficulties arise in either scenario.   A core-degenerate scenario has also been proposed \citep{sokeretal14}, although this too requires a stellar mass companion. The WD merges with the core of the stellar companion, perhaps an AGB star, and the resultant rapidly rotating WD spins down over time and eventually explodes. 

Recent observations, especially those of the well-studied nearby SN 2011fe, have seemed to favor the double degenerate model \citep{maoz14}.  On the other hand,  signatures of  interaction of the Type Ia SN shock wave with a circumstellar medium, signifying mass-loss from a non-degenerate companion, have also been seen. A famous case is that of time-varying narrow absorption lines of Na I \citep{patatetal07}. A summary of such signatures is given in \citet{lrh23}. Calculations suggest that, in certain scenarios, WDs below the Chandrashekar mass can also explode  \citep{ruiter20}, further muddying the waters. \citet{ts15} have argued that a small fraction of Type Ia SNe may explode inside planetary nebulae. Thus it is likely that multiple channels exist to form Type Ia SNe \citep{lm18, soker19}. In recent years many objects have been found which have a spectrum that mostly resembles that of a Type Ia, but have other properties that are not typical of Type Ia's \citep{taubenberger17}. Consequently, a large number of models have been proposed to explain the formation and properties of Type Ia SNe \citep{rs18,tn19}.

{\bf Type Ia SNe with a Hydrogen Line in the Optical Spectrum:} In the scenario where the companion is a main-sequence or evolved star,  H-rich material arising from mass loss by the companion star can accumulate around the progenitor. The SN shock wave expanding outwards can interact with this H-rich medium, thus leading to the presence of H lines in the optical spectrum. In an alternate scenario, the H may be swept-up from the companion star. Either way, these SNe would show the presence of H lines in their optical spectrum. Some Type Ia SNe have been observed to exhibit narrow hydrogen lines superimposed on a SN Ia-like spectrum \citep{hamuyetal03,dengetal04}. The narrow line width implies a velocity lower than that of the typical Type Ia SN shock velocity of around 10,000 km s$^{-1}$ \citep{wangetal09}. In SN 2012ca for example, the line width was varying in time, but the average  H$\alpha$ blue-side width at zero intensity was about 3200 km s$^{-1}$. A possible explanation for this is that the line arises from the SN shock expansion into a dense surrounding medium, which slows the shock, and the width reflects the lower velocity. These SNe (currently $\sim$ 30 in number) have been grouped together under the subclass of Type Ia-CSM \citep{silvermanetal13, sharmaetal23}. They have a distinct H line, with H$\alpha$ luminosities of order 10$^{40}$ to 10$^{41}$ erg s$^{-1}$, Balmer decrements (ratio of H$\alpha$ to H$\beta$ intensity) higher than the nominal ratio of 3, and larger absolute magnitudes compared to normal Type Ia's.

Three low luminosity Type Ia SNe that show the presence of an H$\alpha$ line in their optical spectrum have also been detected in recent years. These SNe, SN 2018fhw \citep{kollmeieretal19, vallelyetal19}, SN 2016jae \citep{eliasrosaetal21}, and SN 2018cqj \citep{prietoetal20}, do not share the other characteristics that are typical of Type Ia-CSM SNe. They are subluminous Type Ia's, generally occur in early type galaxies with older stellar populations, and their H$\alpha$ line luminosity is about two orders of magnitude lower than in the Ia-CSMs. They all appear to be bona-fide Type Ia SNe that show an H$\alpha$ line after several weeks/months. Whether these SNe form low-luminosity members of the Ia-CSM sub-class, or whether they comprise a separate sub-class, is not apparent. The observed H$\alpha$ line suggests some similarity in physical characteristics. On the other hand, there is no question that SN 2018fhw, SN 2018cqj and SN 2016jae are definite Type Ia SNe, whereas doubts about the Type Ia origin of the Ia-CSMs still persist. \cite{kollmeieretal19} had suggested that up to 10\% of sub-luminous fast-declining SNe could show evidence of H$\alpha$ at nebular phases. The discovery of three such SNe suggests that there may be others. 

An over-luminous Type Ia, SN 2015cp, also had narrow H$\alpha$ detected nearly 700 days after explosion \citep{grahametal19}. The H$\alpha$ line in this SN appeared to arise much later than in other Type Ia SNe that showed H lines, and had faded by about 800 days. 

The presence of a H line in the optical spectrum of a Type Ia requires interaction with a  H-rich medium. In general this must be a significantly dense surrounding medium (\S \ref{sec:disc}), needed to give rise to the H$\alpha$ emission, as was shown for SN 2012ca \citep{bocheneketal18}, and discussed in the case of SN 2018fhw \citep{dwarkadas23a}.  Interaction of the SN shock wave with the surrounding medium has generally been studied at X-ray and radio wavelengths, but has been seen mainly in core-collapse SNe. Over 200 radio SNe have been detected \citep{bietenholzetal21}. While no Type Ia's with a H line have been detected in the radio, recently a Type Ia with a He-line in its optical spectrum has been detected at radio wavelengths \citep{kooletal23}. Over 65 SNe have been detected in X-rays \citep{dg12, vvd14, drrb16, rd17, bocheneketal18}, with upper limits for many others. All but one have been of the core-collapse variety. The singular exception is SN 2012ca,  which represents the first and only detection of X-ray emission from a Type Ia SN \citep{bocheneketal18}. 

Other searches for X-ray emission from Type Ia's \citep{sp93, marguttietal14, sandetal21} have failed to detect any sources besides SN 2012ca. The relatively nearby Type Ia SN 2011fe was also not detected in X-rays \citep{horeshetal12}. \citet{hughesetal07} studied 4 Type Ia SNe  using {\it Chandra},  including two, SN 2002ic and 2005gj, that showed signs of interaction and have been classified as Type Ia-CSM SNe \citep{ff13}. They failed to detect X-ray emission from any of the SNe.  \citet{ri12} studied the X-ray emission from 53 Type Ia SNe observed with Swift, but failed to detect X-ray emission from any of them, even in stacked images. SN 2018fhw was imaged with the Chandra ACIS telescope in 2021, but no X-ray emission was detected at the location of the SN \citep{dwarkadas23a}. 

While detection of a Type Ia-CSM in X-rays was highly significant, a sample of one, with a low count rate (especially at the second epoch), does not allow us to comprehend the physics of these objects, decipher their progenitors, or investigate their environment. Arguments against SN 2012ca being a Type Ia SN \citep{inserraetal14} were effectively contradicted by  \citet{foxetal15}, but questions about their origin still persist. In order to establish Type Ia SNe as a new class of X-ray SNe, it is imperative that we find, and detect with higher significance, more Type 1a SNe in X-rays. In the face of mounting evidence that Ia's arise from double degenerate systems, the detection of a Type 1a SN in X-rays could indicate the presence of a  dense surrounding medium, arising from a companion star or being stripped off the companion, thus signifying a single degenerate progenitor.  

In this paper we continue our X-ray investigations of Type Ia SNe that show a H line in the optical spectrum. In \S 2 we report on Chandra observations of the low-luminosity Type Ia SN 2016jae. \S 3 expands on our initial brief study \citep{dwarkadas23b} of the low-luminosity Type Ia SN2018cqj/ATLAS18qtd (hereafter referred to as SN 2018cqj).   In \S \ref{sec:comp} we compare the upper limits for all Type Ia-CSMs and low-luminosity Type Ia that have been observed in X-rays. \S \ref{sec:disc} estimates the density of the medium around the SN at the time that the H$\alpha$ line was detected, and the resultant X-ray luminosity, and then discusses parameters at the time of the Chandra observation. Finally \S \ref{sec:results} summarizes our work and emphasizes the importance of continual observations in the X-ray regime.

\section{SN 2016jae}
\label{sec:16jae}

\begin{table}
	\centering
	\caption{Table of SN 2016jae X-Ray Observations}
	\label{tab:obs16jae}
	\begin{tabular}{lcc} 
		\hline
		Name & ObsID & Exposure \\
		            &   &  (ks)         \\
		 \hline
		SN 2016jae & 26612 &   13.79 \\
		SN 2016jae & 27686 &   9.76 \\
		SN 2016jae & 27687 &   17.84 \\
		SN 2016jae & 27688 &   10.17 \\
		\hline
	\end{tabular}
\end{table}

SN 2016jae was discovered on 2016 December 21.99 UT by the MASTER Global Robotic Net \citep{gressetal16}, with an unfiltered magnitude of 17.2. It was  independently discovered by the Asteroid Terrestrial-impact Last Alert System \citep[ATLAS;][]{tonryetal18}, the Gaia transient survey \citep{hodgkinetal13}, and Pan-STARRS \citep{chambersetal16, magnieretal20}. Classification as a SN1a was done \citep{smithetal16a,smithetal16b} by the Public ESO Spectroscopic Survey for Transient Objects (PESSTO).

\citet{eliasrosaetal21} adopted a redshift for the SN of 0.021, giving a luminosity distance of 92.9 $\pm$ 4.3 Mpc. Their spectra showed the presence of an H$\alpha$ line at two epochs, 84 and 142 days after peak. The line was weak at 84 days, with FWHM 650 km s$^{-1}$, and a line luminosity 3 $\pm\, 0.8 \times 10^{38}$ erg s$^{-1}$. It was distinctly visible at the second epoch, with FWHM $\sim$ 1000 km s$^{-1}$, and a luminosity 1.6 $\pm\, 0.2 \times 10^{38}$ erg s$^{-1}$. The line luminosity had decreased by about a factor of two between the epochs, although given the error bars it is possible that the luminosity may have been more or less constant. A 10$\sigma$ upper limit on H$\beta$ of $1\times 10^{37}$ erg s$^{-1}$ was obtained by the authors, thus suggesting a ratio of H$\alpha$ to H$\beta$ flux $ \ge 16$. This is not inconsistent with SN 2018fhw as well as SN 2018cqj.

\begin{figure*}
\includegraphics[width=1.05\textwidth]{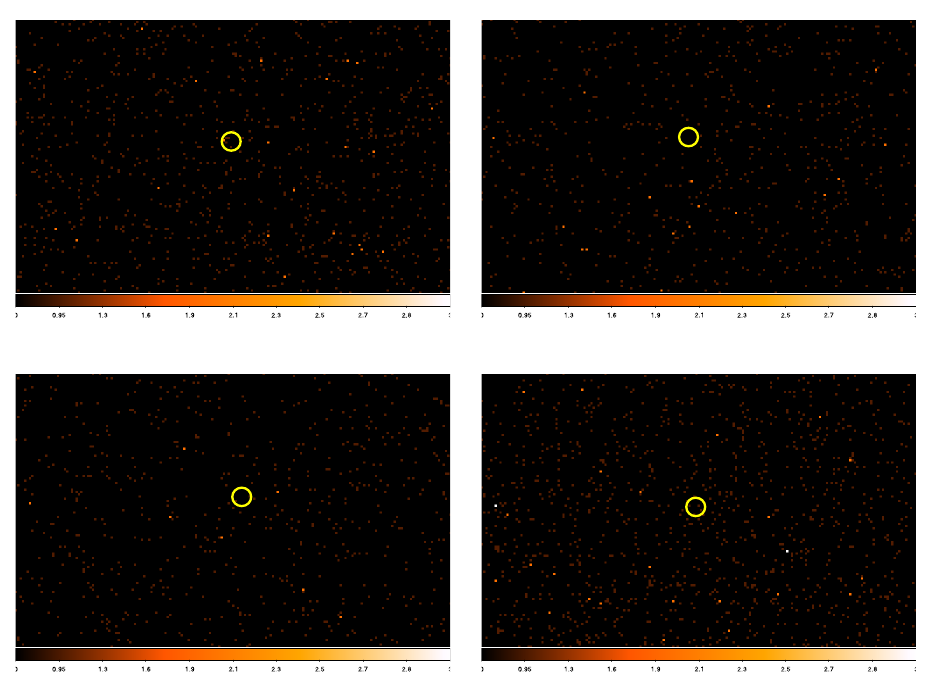}
    \caption{ACIS-S images of the region containing SN 2016jae. The SN position is marked with a yellow circle of radius 2$^{\prime\prime}$ in each case. [Top Left] ObsID 26612. [Top Right] ObsID 27686. [Bottom Left] ObsID 27688. [Bottom Right]. ObsID 27687. No X-ray emission within 0.5-8 keV is detected at the position of the SN in any of the datasets.}
    \label{fig:2016jae}
\end{figure*}

The presence of the H$\alpha$ line suggested the possibility that the SN was expanding into a  dense medium (\S \ref{sec:disc}), and further that this interaction could be observed in X-rays. We therefore proposed a Chandra observation of SN 2016jae. The proposal was accepted, and  SN 2016jae was observed with Chandra ACIS-S in early February 2023, approximately 2233 d after discovery, or 2187.5 d in the rest frame. The $\sim$50 ks observations was scheduled as 4 separate observations of 10-18 ks each, as detailed in Table~\ref{tab:obs16jae}. All data were downloaded and analyzed using Chandra CIAO 4.15 and CalDB 4.10.4.  We inspected and reduced each of the four datasets separately. No emission was detected at the position of SN 2016jae in any of the datasets (Figure~\ref{fig:2016jae}). Using the CIAO command {\tt specextract}, spectra were extracted from each of the four datasets. For each of the datasets we used a 2$^{\prime\prime}$ source region. In two of the datasets (ObsIDs 26612 and 27686), the background region was an annulus with a 2$^{\prime\prime}$ inner radius and an outer radius of 5$^{\prime\prime}$ around the SN.  For the other two datasets (ObsIDs 27687 and 27688), the same background region gave an error as it had zero counts in the 0.2-3 keV range needed to create a weights map. The error persisted even when the outer radius was increased to 8$^{\prime\prime}$. Therefore in these two cases we used an 8$^{\prime\prime}$ background region centered on a different location close to the SN. Having created individual spectra for each dataset, we then combined all the spectra using the {\tt combine\_spectra} script in CIAO. The combined spectrum had 0 counts in the source region, and 8.33 counts in the background region. The ratio of source to background areas was 0.12. We use the Bayesian method of \citet{kbn91} to determine the maximum number of counts in the region, and derive a 99.7\% (3$\sigma$) upper limit of 5.795 counts in 51,574 s. This gives a maximum count rate of 1.12 $\times 10^{-4}$ c s$^{-1}$. This count rate is used in Chandra PIMMS\footnote{https://cxc.harvard.edu/toolkit/pimms.jsp} to calculate the flux assuming different emission models. 

In order to calculate the upper limit to the flux, we follow the same procedure as \citet{dwarkadas23a}. Although SN 2016jae differs from SNe of Type Ia-CSM in that it has lower  H$\alpha$ luminosity, and is found in a low luminosity galaxy, the presence of a detectable H$\alpha$ line suggests the presence of a high density medium around the SN. The high density can result in a slowing down of the shock wave. If the density is high, the most likely emission mechanism in these SNe would then be thermal emission, due to thermal bremsstrahlung combined with line emission. Thermal emission depends on the square of the density of the emitting plasma, and therefore of the density of the surrounding medium, and will be higher at high densities. It was postulated as the X-ray emission for SN 2012ca by \citet{bocheneketal18}. Although we take this to be the preferred model for the emission, we also quote a flux using a power-law model, which may be more suitable if the density has substantially decreased, and Inverse Compton emission or synchrotron is the most likely source of any emission. 

The presence of a high density medium near the star will lead to a high column density around the SN, which must be taken into account in addition to the Galactic column towards the source, in order to calculate the intrinsic luminosity, as done in \citet{hughesetal07}.  In SN 2012ca, \citet{bocheneketal18} found a column density likely exceeding 1 $\times 10^{22}$ cm$^{-2}$ cm$^{-2}$ for a temperature of a few keV. Without knowledge of the value of the density of the medium, it is difficult to predict the column density.   A mass-loss rate around 1 $\times \,10^{-4}$ M$_{\odot}$ yr$^{-1}$ will result in a column density of 10$^{23}$ cm$^{-2}$. A higher mass-loss rate can result in an even higher column density. We note that the column density calculation is accompanied by several caveats. Firstly, as we discuss later, the density, and therefore the parameters of any potential wind, must be changing with time. Additionally, the freely expanding wind cannot interact directly with the ambient medium; the interaction of a freely expanding wind with the surrounding medium must result in the formation of a wind bubble \citep{weaveretal77}, which complicates the column density calculation. Finally, a higher wind density will lead to more X-ray emission, which depends on the density squared. The X-rays themselves can ionize the surrounding medium, effectively reducing the column density. To take into account the variation in mass-loss rates, we assume a range of intrinsic column densities, from 1 $\times 10^{21}$ cm$^{-2}$ cm$^{-2}$ to 1 $\times 10^{23}$ cm$^{-2}$. These account for material surrounding the SN, as well as any column density in the host galaxy in the direction of the SN.  The flux is determined assuming this additional redshifted column along with the Galactic one. We use the PLASMA/APEC model in PIMMS to calculate the thermal emission, corresponding to thermal bremsstrahlung combined with line emission.  A  range of X-ray temperatures, from 1.53 keV to 9.67 keV is considered.  Solar abundances are used, and 3$\sigma$ results are quoted. For the power-law model, a power-law index of 2, and a Galactic column density is used. With the help of Colden\footnote{https://cxc.harvard.edu/toolkit/colden.jsp}, we estimate a Galactic column density towards the source of 3.09 $\times 10^{20}$ cm$^{-2}$ \citep{dl90}.  Our results for SN 2016jae, quoted in the 0.5-8 keV band, are given in table~\ref{tab:flux16jae}. The luminosities all lie below about 2. $\times10^{40}$ erg s$^{-1}$.

\begin{table}
	\centering
	\caption{The unabsorbed flux and luminosity of SN 2016jae, assuming various emission models with different parameters, and a distance of 92.9 Mpc. `Temp'=Temperature of the plasma. Column refers to the intrinsic column, in addition to the Galactic column of 3.09 $\times 10^{20}$ towards the source. 3$\sigma$ results are quoted in the 0.5-8 keV band. }
	\label{tab:flux16jae}
	\begin{tabular}{lcccc} 
		\hline
		Model & Temp & Column & Flux & Luminosity\\
		            & (keV)   &  (cm$^{-2}$)          & (erg s$^{-1}$ cm$^{-2}) $ & (erg s$^{-1}$) \\
		\hline
		Power-law, $\Gamma$=2 &  &  & 1.77 $\times 10^{-15}$ & 1.8 $\times 10^{39}$ \\
		 \hline
		Plasma/APEC & 1.93  &  &  1.44 $\times 10^{-15}$ & 1.5 $\times 10^{39}$ \\
		Plasma/APEC & 4.85   &   &  1.7 $\times 10^{-15}$ & 1.75 $\times 10^{39}$ \\
		Plasma/APEC & 9.67  &   &  1.82 $\times 10^{-15}$ & 1.9 $\times 10^{39}$ \\
		\hline
		Plasma/APEC & 1.53  & 1. $\times 10^{21}$   & 1. 5 $\times 10^{-15}$ & 1.6 $\times 10^{39}$ \\
		Plasma/APEC & 1.53 & 5. $\times 10^{21}$  & 2.0 $\times 10^{-15}$ & 2.1 $\times 10^{39}$ \\
		Plasma/APEC & 1.53 &  1. $\times 10^{22}$ & 2.7 $\times 10^{-15}$ & 2.8 $\times 10^{39}$ \\
		Plasma/APEC & 1.53 &  2. $\times 10^{22}$ & 4.1 $\times 10^{-15}$ & 4.3 $\times 10^{39}$ \\
		Plasma/APEC & 1.53 &  1. $\times 10^{23}$ & 1.9 $\times 10^{-14}$ & 2. $\times 10^{40}$ \\
		\hline
		Plasma/APEC & 3.06  & 1. $\times 10^{21}$   & 1.6 $\times 10^{-15}$ & 1.7 $\times 10^{39}$ \\
		Plasma/APEC & 3.06  & 5. $\times 10^{21}$  & 2. $\times 10^{-15}$ & 2.05 $\times 10^{39}$ \\
		Plasma/APEC & 3.06  &  1. $\times 10^{22}$ & 2.4 $\times 10^{-15}$ & 2.5 $\times 10^{39}$ \\
		Plasma/APEC & 3.06  &  2. $\times 10^{22}$ & 3.2 $\times 10^{-15}$ & 3.35 $\times 10^{39}$ \\
		Plasma/APEC & 3.06  &  1. $\times 10^{23}$ & 9.4 $\times 10^{-15}$ & 9.7 $\times 10^{39}$ \\
		\hline
		Plasma/APEC & 4.85  & 1. $\times 10^{21}$  & 1.8 $\times 10^{-15}$ & 1.8 $\times 10^{39}$ \\
		Plasma/APEC & 4.85  & 5. $\times 10^{21}$  & 2.1 $\times 10^{-15}$ & 2.2 $\times 10^{39}$ \\
		Plasma/APEC & 4.85  & 1. $\times 10^{22}$  & 2.5 $\times 10^{-15}$ & 2.5 $\times 10^{39}$ \\
		Plasma/APEC & 4.85  & 2. $\times 10^{22}$  & 3.2 $\times 10^{-15}$ & 3.3 $\times 10^{39}$ \\
		Plasma/APEC & 4.85  & 1. $\times 10^{23}$  & 8.0 $\times 10^{-15}$ & 8.2 $\times 10^{39}$ \\
		\hline
	\end{tabular}
\end{table}

\section{SN 2018cqj}
\label{sec:18cqj}

SN 2018cqj/ATLAS18qtd (hereafter SN 2018cqj) was discovered by the Asteroid Terrestrial-impact Last Alert System (ATLAS) transient survey on 2018 June 13.27. The co-ordinates were RA=09:40:21.463 and DEC = -06:59:19.76 (J2000.0). Although no obvious host galaxy was found at this location, the NED database showed that the transient was located 3.1$^{\prime}$ from the nearby S0 galaxy IC550. Since the redshift of the SN appeared to be consistent with that of the galaxy, \citet{prietoetal20} suggested that the SN was physically associated with this galaxy. This gave a redshift of 0.0165  for the SN, and a luminosity distance of 74.3 Mpc, assuming a Hubble constant of 72 km s$^{-1}$ Mpc$^{-1}$. The V-band decline rate of SN2018cqj at 200-300 days after peak was
consistent with the decline rates of SN2011fe and SN1991bg at the same late-time epochs, thus confirming that it was a typical Tye Ia SN in that respect.

SN 2018cqj also showed the presence of an H$\alpha$ line 193 and 307 days after peak. The line was resolved at the first epoch to have a FWHM of 1200 km s$^{-1}$ and a luminosity of $\approx 3.8 \pm 0.9 \times 10^{37}$ erg s$^{-1}$. The line was unresolved at the 2nd epoch, and its luminosity decreased by almost an order of magnitude to  $\approx 4.6 \pm 1.4 \times 10^{36}$ erg s$^{-1}$. \citet{prietoetal20} also searched for an H$\beta$ line, but did not detect it, finding a lower limit to the Balmer decrement $F(H\alpha) / F (H\beta) \ge 6$, consistent with that of SN 2018fhw.

We observed the SN with the ACIS-S instrument on the Chandra satellite on 2022 December 12 (ObsID 26613, PI Dwarkadas), 1643 days after discovery (1616 in the SN rest frame). The observation consisted of a single exposure of 44.88 ks. The data were downloaded and analysed as for SN 2016jae, using CIAO and Sherpa.

\begin{figure*}
	\includegraphics[width=0.9\textwidth]{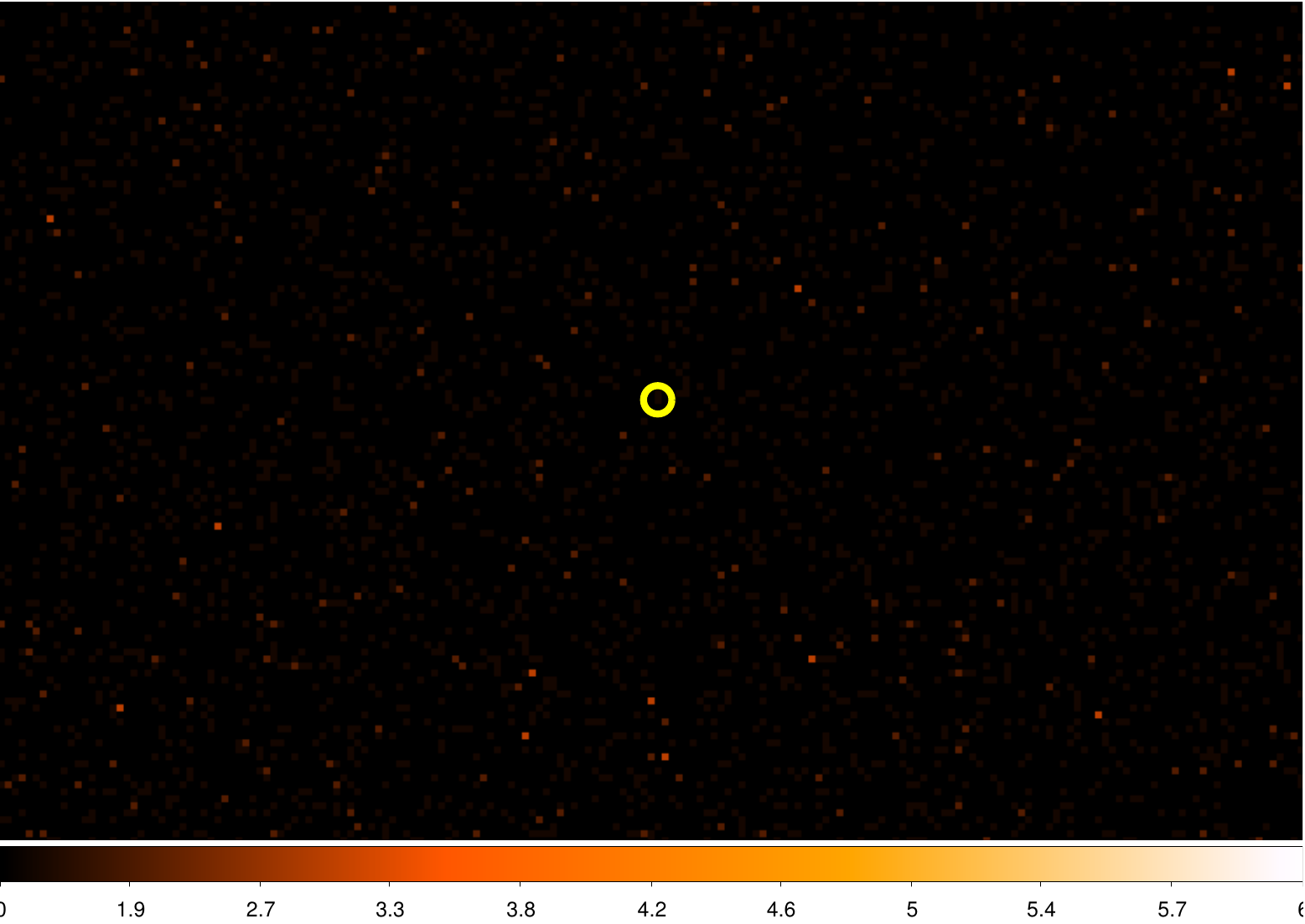}
    \caption{ACIS-S image of the region containing SN 2018cqj. The SN position is marked with a yellow circle of radius 1$^{\prime\prime}$. No X-ray emission is detected at that position.}
    \label{fig:18cqj}
\end{figure*}

Figure~\ref{fig:18cqj} shows the ACIS-S image of the region containing SN 2018cqj. The Simbad position is marked with a yellow circle. No X-ray emission is detected at this position.

In order to calculate upper limits for SN 2018cqj, we use a source region of 1$^{\prime\prime}$ around the SN, which includes one count within the range 0.5-8 keV. The background region is an annulus with inner radius 1$^{\prime\prime}$ and outer radius 5$^{\prime\prime}$, which contains 8 counts in the 0.5-8 keV X-ray range. This gives 0.33 counts in the source region. We again use the Bayesian method of \citet{kbn91} to determine the maximum number of counts in the region, and derive a 99.7\% (3$\sigma$) upper limit of 7.592 counts in 44,879 s. This results in a count rate of 1.69 $\times 10^{-4}$ counts s$^{-1}$. This rate is used in PIMMS to get the flux. A Galactic column density towards the source of 3.41 $\times\, 10^{20}$ cm$^{-2}$ \citep{dl90} is obtained from Colden. A redshift of 0.017, and distance to the source of 74.3 Mpc are adopted. Our values for the flux and the resultant luminosity, calculated from the source counts using Chandra PIMMS, are given in Table \ref{tab:flux18cqj}. For additional confirmation, we also used the Chandra CIAO routine {\em srcflux} to calculate the 3$\sigma$ flux within 0.5-8 keV. This routine returned a maximum flux of 2.9 $\times 10^{-15}$ erg s$^{-1}$ cm$^{-2}$ (3$\sigma$ value) for a power-law model with an index of 2,  consistent with the values determined using the method of \citet{kbn91}. For SN 2018cqj also we find 3$\sigma$ instrinsic  luminosities lying generally below about 2 $\times10^{40}$ erg s$^{-1}$.

\begin{table}
	\centering
	\caption{The unabsorbed flux and luminosity of SN 2018cqj, assuming various emission models with different parameters, and a distance of 74.3 Mpc. `Temp'=Temperature of the plasma. Column refers to the intrinsic column, in addition to the Galactic column of 3.41 $\times 10^{20}$ towards the source. 3$\sigma$ results are quoted.}
	\label{tab:flux18cqj}
	\begin{tabular}{lcccc} 
		\hline
		Model & Temp & Column & Flux & Luminosity\\
		            & (keV)   &  (cm$^{-2}$)          & (erg s$^{-1}$ cm$^{-2}) $ & (erg s$^{-1}$) \\
		\hline
		Power-law, $\Gamma$=2 &  &  & 2.67 $\times 10^{-15}$ & 1.8 $\times 10^{39}$ \\
		 (Using {\it srcflux})  &  &  &  2.9 $\times 10^{-15}$ & 1.9 $\times 10^{39}$ \\
		 \hline
		Plasma/APEC & 1.93  &  &  2.18 $\times 10^{-15}$ & 1.4 $\times 10^{39}$ \\
		Plasma/APEC & 4.85   &   &  2.56 $\times 10^{-15}$ & 1. 7$\times 10^{39}$ \\
		Plasma/APEC & 9.67  &   &  2.75 $\times 10^{-15}$ & 1.8 $\times 10^{39}$ \\
		\hline
		Plasma/APEC & 1.53  & 1. $\times 10^{21}$   &  2.35 $\times 10^{-15}$ & 1.55 $\times 10^{39}$ \\
		Plasma/APEC & 1.53  & 5. $\times 10^{21}$  & 3.1 $\times 10^{-15}$ & 2. $\times 10^{39}$ \\
		Plasma/APEC & 1.53  &  1. $\times 10^{22}$ & 4.1$\times 10^{-15}$ & 2.7 $\times 10^{39}$ \\
		Plasma/APEC & 1.53  &  2. $\times 10^{22}$ & 6.3 $\times 10^{-15}$ & 4.1 $\times 10^{39}$ \\
		Plasma/APEC & 1.53  &  1. $\times 10^{23}$ & 2.9 $\times 10^{-14}$ & 1.9 $\times 10^{40}$ \\
		\hline
		Plasma/APEC & 3.06  & 1. $\times 10^{21}$   &  2.5 $\times 10^{-15}$ & 1.6 $\times 10^{39}$ \\
		Plasma/APEC & 3.06  & 5. $\times 10^{21}$  & 3. $\times 10^{-15}$ & 2. $\times 10^{39}$ \\
		Plasma/APEC & 3.06  &  1. $\times 10^{22}$ & 3.7 $\times 10^{-15}$ & 2.4 $\times 10^{39}$ \\
		Plasma/APEC & 3.06  & 2. $\times 10^{22}$  & 4.9 $\times 10^{-15}$ & 3.2 $\times 10^{39}$ \\
		Plasma/APEC & 3.06  & 1. $\times 10^{23}$  & 1.4 $\times 10^{-14}$ & 9.2 $\times 10^{39}$ \\
		\hline
		Plasma/APEC & 4.85  & 1. $\times 10^{21}$  & 2.7 $\times 10^{-15}$ & 1.8 $\times 10^{39}$ \\
		Plasma/APEC & 4.85  & 5. $\times 10^{21}$  &3.2 $\times 10^{-15}$ & 2.1 $\times 10^{39}$ \\
		Plasma/APEC & 4.85  & 1. $\times 10^{22}$  & 3.7 $\times 10^{-15}$ & 2.4 $\times 10^{39}$ \\
		Plasma/APEC & 4.85  & 2. $\times 10^{22}$  & 4.8 $\times 10^{-15}$ & 3.2 $\times 10^{39}$ \\
		Plasma/APEC & 4.85  & 1. $\times 10^{23}$  & 1.2 $\times 10^{-14}$ & 7.9 $\times 10^{39}$ \\
		\hline
	\end{tabular}
\end{table}

\section{Comparison to other SNe} 
\label{sec:comp}

So far 3 low optical luminosity Type Ia SNe with an H$\alpha$ line in their optical spectrum have been imaged in X-rays with Chandra: SN 2018fhw, SN 2018cqj and SN 2016jae. No X-ray emission was  detected in any of these SNe. These SNe are somewhat different from the Type Ia-CSM SNe, some of which have also been observed in X-rays with Chandra \citep{hughesetal07}. Amongst all these SNe, as well as other Type Ia SNe imaged in X-rays, SN 2012ca remains the only Type Ia SN in which X-ray emission has been detected. SN 2012ca was observed twice with the {\it Chandra} X-ray telescope. The first X-ray observation of SN 2012ca, at 554 days past explosion, resulted in around 30 counts and a solid detection. The second at 745 days yielded 10 counts. The measured Balmer decrement  ranged from 3-20. The most conservative estimate of the surrounding density yielded a value of at least 10$^6$ particles cm$^{-3}$. The authors preferred a two-component medium, with the X-ray emitting material having a much higher density, around 10$^8$ cm$^{-3}$. In either case, it was clear that in order to reach the observed X-ray intensity, the SN must have been surrounded by a dense medium with which the shock interacted to produce the observed X-rays.

\begin{figure}
	\includegraphics[width=\columnwidth]{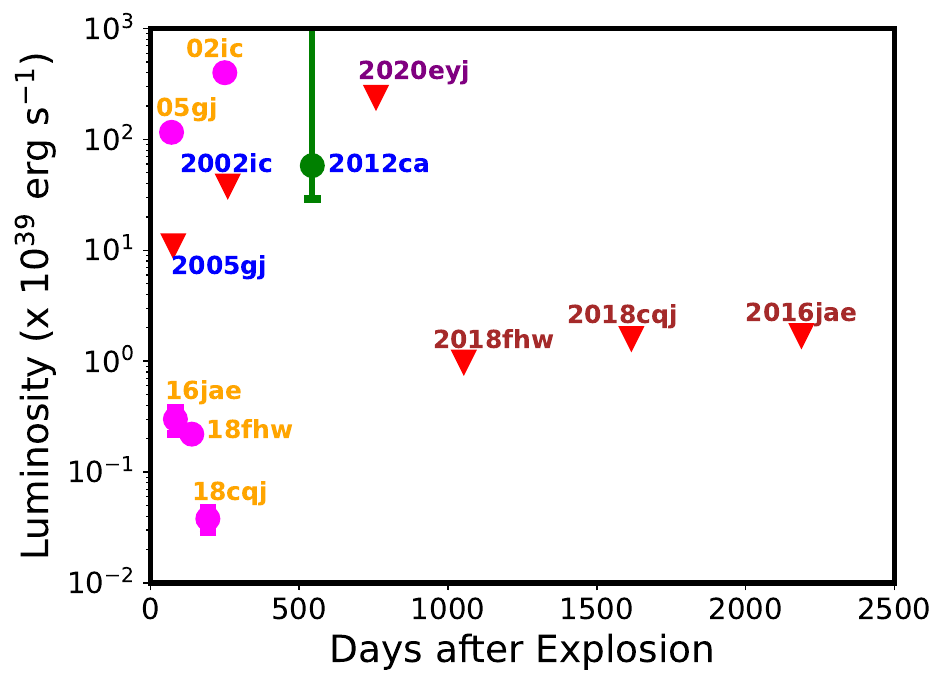}
	\caption{Comparison of the observed X-ray luminosity of SN 2012ca with the upper limits for all other Type Ia SNe which show a H line in their optical spectrum. X-Ray upper limits are shown for a temperature of 3.06 keV and a column density of 1~$\times\, 10^{21}$ cm$^{-2}$. The Type Ia-CSM SNe are labelled in blue, the low-luminosity Ia's are labelled in brown, and 2020eyj, seen in radio but not in X-rays, is labelled in purple. The observed H$\alpha$ luminosity for each SN (except 2020eyj) is also shown, generally at the epoch at which it was maximum, and labelled in orange. X-ray observations of Type Ia CSM SNe 2005gj and 2002ic occurred close to the time when H$\alpha$ emission was observed. In the case of the three low luminosity Type Ia SNe, the X-ray observations were all taken years after the H$\alpha$ observations. For SN 2012ca the H$\alpha$ luminosity is not shown. Although H$\alpha$ observations were taken at several epochs, the luminosities are not available in the literature. However H$\alpha$ emission was seen beyond 500 days, close to the epoch of the first X-ray observations.
    	\label{fig:snxray}}
\end{figure}

In Figure~\ref{fig:snxray} we show the results from X-ray observations of all the Type Ia SNe with a H line in their optical spectrum. For the sake of comparison, we plot all upper limits assuming a temperature of 3.06 keV and a column density of 1~$\times\, 10^{21}$ cm$^{-2}$ in addition to the Galactic column towards the source. Only the first epoch of detection of SN 2012ca is shown. The data for SN 2012ca is from \citet{bocheneketal18}, and for SN 2018fhw, SN 2005gj and SN 2002ic from \citet{dwarkadas23a}. For completeness we also show the upper limit from SN 202eyj, the only Type Ia SN to be detected at radio wavelengths \citep{kooletal23}. This did not show a H line in its optical spectrum, but it did have a He-rich circumstellar medium. Unfortunately it was only observed for 3.8 ks  in X-rays with the Swift-XRT telescope. The upper limit is too high to provide meaningful constraints.

The low optical luminosity Type Ia's all have X-ray upper limits an order of magnitude or so lower than the Type Ia-CSMs. This could be because they are intrinsically lower luminosity, although we cannot exclude the fact that they have all been observed in X-rays at much later times ($>$ 1000 d) compared to the Ia-CSMs. The only 1a SN that has been detected in X-rays, 2012ca, was also highly luminous, which would have contributed to its detectability.  None of the other SNe have been observed in X-rays at the same epoch as 2012ca. The other Type Ia SNe showing a  H line had X-ray observations taken either about a year earlier than SN 2012ca, or at least a year later. SN 2020eyj was observed at 758 d, similar to the second detection of SN 2012ca at 754 d, albeit with a very short exposure using the Neil Gehrels Swift Observatory. Given its luminosity at 745 d, SN 2012ca would not have been detected in a 4 ks Swift observation.

In the low luminosity Type Ia SNe, as well as in SN 2005gj and SN 2002ic, the presence of an H$\alpha$ line in the optical spectrum is noted only in the first year after explosion. In 2020eyj it is not detected at all, although He emission lines are detected. Only in SN 2012ca was the H$\alpha$ line visible past 550 days. This may indicate that in SN 2012ca the circumstellar interaction continued on for at least 1.5 years, and thus the density was high at least up until the time of the first X-ray observation, contributing to its detectability.

Figure~\ref{fig:snxray} also includes the H$\alpha$ luminosity (pink) when it is available in the literature. Where H$\alpha$ was seen at multiple epochs, the luminosity is plotted at the epoch at which it was maximum. In most cases this is less than one year after explosion. In many Type Ia SNe there were multiple epochs where the H$\alpha$ was observed. In the case of the low luminosity 1a's, these were all generally within the first year, whereas the X-ray observations happened a few years later. In the case of SN 2002ic and SN 2005gj, the X-ray data were taken close to the epoch of maximum H$\alpha$. In both those cases, the X-ray upper limits (for a column density of 1~$\times\, 10^{21}$ cm$^{-2}$) were lower than the H$\alpha$ luminosity. This suggests that the X-ray emission may not have been responsible for ionizing the H$\alpha$, unless the X-ray luminosity was at least two orders of magnitude higher, which implies a substantially higher column density. In the case of SN 2012ca, the H$\alpha$ line was detected past 500 days, close to the epoch of the first X-ray observation at 554 days. Unfortunately the line luminosities are not available in the literature, but it appears that the the density was still quite high, although perhaps not as high as at the epoch of maximum H$\alpha$. In the case of the low-luminosity Type Ia SNe, the X-ray upper limits are a few times to an order of magnitude higher than the H$\alpha$ luminosity, suggesting that X-rays could have been responsible for ionizing the hydrogen. However, the time delay between the epoch of H$\alpha$ and X-ray observations lends some doubt to this conclusion.

\begin{figure}
	\includegraphics[width=\columnwidth]{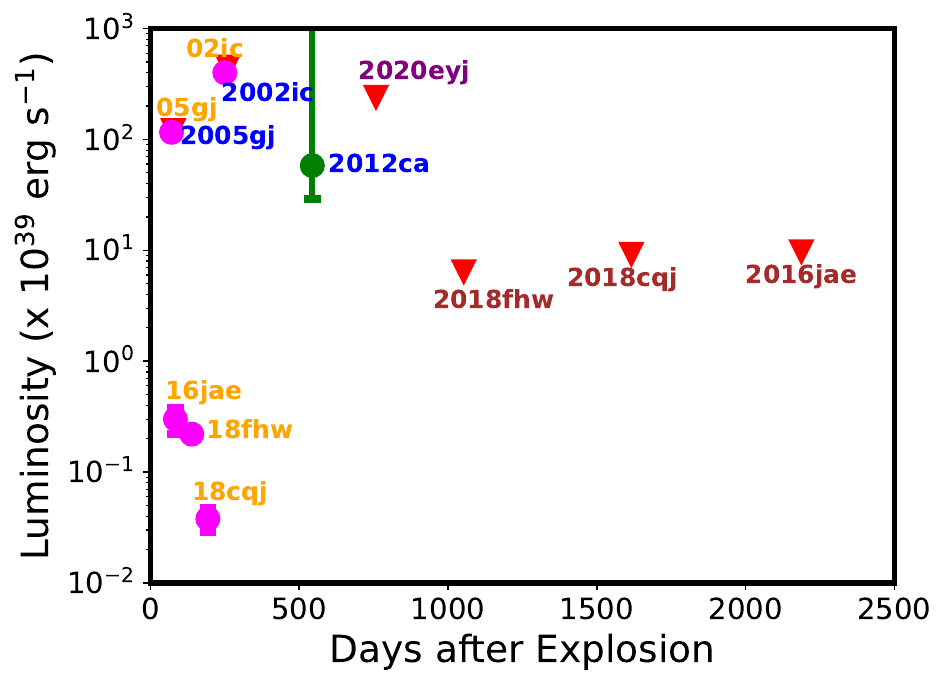}
    \caption{Similar to Figure ~\ref{fig:snxray}, except that X-ray upper limits are shown for an X-ray temperature of 3.06 keV and a column density of 1~$\times\, 10^{23}$ cm$^{-2}$.
    	\label{fig:snxray2}}
\end{figure}

Figure~\ref{fig:snxray2} is similar to Figure~\ref{fig:snxray}, with the exception that the column density has been increased to 1~$\times\, 10^{23}$ cm$^{-2}$ to calculate the X-ray upper limits. The X-ray upper limits for SN 2018fhw, SN 2002ic and SN 2005gj from \citet{dwarkadas23a} were recalculated for this column density, using the parameters given in that paper.  Even though the column density is 100 times higher than that used in Figure~\ref{fig:snxray}, it can be seen that the inferred X-ray upper limit is still just about equal to the maximum observed H$\alpha$ luminosity for the Type Ia-CSM SNe.  \citet{cf94} find that the H$\alpha$ emission must be about 5\% of the X-ray flux if the X-rays are ionizing the H$\alpha$. This may suggest that the X-rays are insufficient to ionize the H$\alpha$ in the Type Ia CSM SNe, except for column densities significantly in excess of $10^{23}$ cm$^{-2}$, which are rarely seen in practice. This is not the case for the low luminosity Ia SNe, where the X-ray upper limits are substantially higher than the H$\alpha$ luminosity. It is possible that the ionization process for H$\alpha$ is different in the low-luminosity Type Ia SNe as compared to the Type Ia-CSM SNe.

\section{Discussion}
\label{sec:disc}

{\bf The H$\alpha$ line:} The presence of the H$\alpha$ line in the optical spectrum is the main clue pointing to a dense medium around these SNe compared to other Type Ia's.   Analogous to SN 2012ca, we expect X-ray emission to be due to thermal bremsstrahlung, suggested by the low measured velocity and the high density indicated by the H$\alpha$ line (below).  The FWHM of the H$\alpha$ lines is $\approx 1000 $ km s$^{-1}$. This could be considered a proxy for the shock velocity, although other inferences are possible.  The magnitude of thermal X-ray emission (proportional to the density squared) depends crucially on the density of the surrounding medium, and therefore on the mechanism that gives rise to the H$\alpha$, which is unclear. In SN 2012ca the high luminosity of the H$\alpha$ line, observed starting 50 days from the time of explosion, and the high Balmer decrement (3-20), was interpreted as being due to collisional excitation of the line.

The H$\alpha$ line was seen at two epochs in both SN 2016jae and SN 2018cqj, while the H$\beta$ line was not visible in either. We investigate various processes that could give rise to the line, and the corresponding density estimated in each case. \citet{dwarkadas23a} showed that the effects of electron scattering were unlikely to be important for SN 2018fhw.  Given the much later time after explosion at which SN 2018cqj and 2016jae were imaged, and the shape of the line, which was fitted by multiple gaussians in either case, the same arguments hold true for SN 2018cqj and SN 2016jae. We do not consider electron scattering as a viable possibility in these two SNe. The Balmer decrement in SN 2018cqj  has a value $>$ 6,  while in SN 2016jae it could be higher. Although this would in general preclude case B recombination, given the uncertainty in measurements, and the probability of selective dust extinction, we include it as a possibility. For each H$\alpha$ mechanism mentioned below, we compute the density required for the observed H$\alpha$ luminosity, following the procedure outlined in \citet{dwarkadas23a}. 

\begin{itemize}
\item{\em Case B recombination:}  The Case B H$\alpha$ luminosity per unit volume is $n_e\,n_p\,\alpha_B h {\nu}_{\alpha}$, with the recombination coefficient $\alpha_B$ = 2.6 $\times 10^{-13}$ cm$^3$ s$^{-1}$ at a temperature of 10$^4$ K. We assume a constant velocity given by the FWHM of the line (probably a lower limit), and that the emission at this stage arises from a thin shell (encased between the forward and reverse shocks) with a shell thickness of about 0.1 \citep{cf94}. The high end of the observed H$\alpha$ luminosity (SN 2016jae at epoch 1) then yields a number density of 1.7$\times 10^9$ particles cm$^{-3}$, whereas at the lower end (SN 2018cqj at epoch 2) it decreases to 1.6 $\times 10^7$ particles cm$^{-3}$.  The average velocity could be higher than the FWHM (SN 2012ca had an average velocity of at least 3200 km s$^{-1}$). An average velocity v$_{sh}$ km s$^{-1}$ would result in a density that varies as (v$_{sh}/{\rm FWHM})^{-3/2}$ cm$^{-3}$. 

\item{\em Radiative shocks:} If the cooling time of the shocked material is smaller than the age of the shock, whose maximum value is the SN age, then the shock wave becomes radiative.  The maximum luminosity from a shock propagating in a region of radius $a$ and density $\rho$ is $2 \pi a^2 \rho v_{sh}^3$. This luminosity must be at least as large as the observed H$\alpha$ luminosity. The density of this region must therefore be $\rho > L_{H\alpha} /(2 \pi a^2 v_{sh}^3$). If we assume that the shock is expanding into a dense, clumpy medium, as for many IIn SNe  \citep{chugai93, cd94} then the radius $a$ refers to the radius of the clumps, and must be smaller than the shock radius. If we take it to be 10\% of the radius, then we get a minimum density of at least 2 $\times 10^9$ particles cm$^{-3}$ (SN 2016jae) or 2 $\times 10^8$ particles cm$^{-3}$ (SN 2018cqj).  If the shock is expanding into a  dense surrounding medium, then the radius $a$ would be the shock radius. In this case the above estimates would decrease by a factor of 100. Therefore at minimum the density must exceed $10^6$ particles cm$^{-3}$. Clearly a high density is needed for the shock to be radiative.
  
\item{\em Balmer Dominated Shocks:} These happen when a fast shock travels into a partially neutral medium. Given the high H$\alpha$ luminosity and the tentative low velocity of about 1000 km s$^{-1}$ assumed for the shock, Balmer dominated shocks will not be able to account for this high luminosity even assuming a neutral fraction of 1 (i.e. totally neutral), which in itself is highly improbable \citep{cr78}.

\item{\em Collisional excitation}  was put forward to explain the very  high Balmer decrements in the case of SN 2012ca. This  requires densities $> 10^8$ particles cm$^{-3}$ \cite{du80}, implying a high-density medium. citet{cummingetal96} find that the H$\alpha$ emission in Type Ia SNe is dominated by collisional excitation in regions where the temperature is a few times 10$^4$ K. 

\item{\em Recombination following ionization by X-ray emission from the shock wave:} The X-ray emission from the shock is probably the main source of ionization of the surrounding medium, which on recombination gives rise to the H$\alpha$. \cite{cf94} showed that in this scenario the H$\alpha$ luminosity must be less than about 5\% of the X-ray luminosity. Thus, the minimum X-ray luminosity must be at least 20 times the H$\alpha$ luminosity, or $> 6 \times 10^{39}$ erg s$^{-1} $(SN 2016jae) to $> 2 \times 10^{37}$ erg s$^{-1} $( for SN 2018cqj at the second epoch).
\end{itemize}

{\bf X-ray luminosity:}  It is clear from the above discussions that a high circumstellar density is needed at the first epoch, but must have decreased by the second epoch, especially in the case of SN 2018cqj. The high circumstellar densities are consistent with detailed calculations of the H$\alpha$ luminosity as a function of the mass-loss rate carried out by \citet{lundqvistetal13}, if one extrapolates to higher luminosities the results presented in Fig 4 of that work. The subsequent decrease in density is consistent with the fact that the H$\alpha$ line in unresolved at the second epoch for SN 2018cqj. 

The X-ray luminosity can be written as L$_x = n_e^2 \,\Lambda$ V where V is the volume of the emitting region, and $\Lambda$ the cooling function. The primary coolant at these temperatures is thermal bremsstrahlung combined with line emission. We assume $\Lambda = 3.5 \times 10^{-23}$. For a density of $ 10^9$ particles cm$^{-3}$, the luminosity is L$_x = 10^{18} \times 3.5 \times 10^{-23}\, $V, which gives L$_x \approx 8.1 \times 10^{40}$ erg s$^{-1}$ for  SN 2016jae at 142 days. The calculated density for SN 2018cqj at 193 d gives L$_x \approx 1.7 \times 10^{39}$ erg s$^{-1}$.  Notably, the FWHM was increasing with time for SN 2016jae. The luminosity for SN 2016jae is higher than the calculated upper limits, suggesting that the luminosity had definitely decreased by the time of X-ray observations. The luminosity for SN 2018cqj is approximately equal to the calculated upper limits, ruling out an increase in the luminosity.  By the second epoch for SN 2018cqj, the X-ray luminosity would likely be lower due to the lower density suggested by the decreased H$\alpha$ luminosity.

It is possible that this high density medium does not exist close in to the star, but starts at a finite radius. In that case the shock velocity would not be constant, but could be much higher at a smaller radius, and then have decreased to the level indicated by the H$\alpha$ linewidth when the shock collided with the high density region. This would lead to a higher X-ray luminosity. Unfortunately, it is difficult to put constraints on such a  scenario without additional information.

If the density were to remain constant up until the time of the X-ray observation, then the X-ray emission will increase in proportion to the volume, which will increase substantially over a period of 4 years. Thus even for a density of 1. $\times 10^7$ cm$^{-3}$, the X-ray luminosity of SN 2016jae  would be greater than 10$^{40}$ erg s$^{-1}$. For a density 10$^9$ cm$^{-3}$ the luminosity would be of order 10$^{44}$ erg s$^{-1}$. We can therefore safely say that the density did not remain constant in SN 2016jae. Similar arguments apply to SN 2018cqj. 

The radius, and hence volume, will be considerably higher after several years.  From \citet{dg12}, equation 2, we can write

\begin{equation}
L_x \propto n_e^2 \, \frac{r}{t} \, r^3  \,\,   \propto n_e^2 \,\frac{r^4}{t}    \propto n_e^2\; t^{4m - 1}
\end{equation}

\noindent
where we have assumed that the SN radius $ r \propto t^m $, $m$ is the expansion parameter of the SN. The value of m for a young SN is generally 0.8-0.9. Thus $4m - 1 \approx 2.2 - 2.6$. Therefore, in order for the luminosity to decrease, $n_e^2$ must decrease faster than $t^{2.2-2.6}$ or approximately faster than $r^3$ by the above assumption. Therefore, the density $n_e$ must decrease faster than $r^{-1.5}$ for the luminosity to decrease, assuming a continuous decrease.  If the density decreased slower than r$^{-1.5}$ the X-ray luminosity could increase. Some increase in the luminosity would be allowed by the upper limits for SN 2018cqj, but not for SN 2016jae. In general however we would expect the luminosity to decrease with time (as it does for almost all other X-ray SNe \citep{dg12}) and thus that the density would be decreasing faster than r$^{-1.5}$. Thus evolution in a  wind with constant parameters, and density decreasing as r$^{-2}$, would be accommodated, although there is no reason to expect a Type Ia SN to be evolving in a constant parameter wind for $> 4$ years.

\section{Conclusions}
\label{sec:results}

In this paper we used the ACIS-S instrument on the Chandra X-ray Observatory to study SN 2018cqj and SN 2016jae. The presence of an H$\alpha$ line in the optical spectrum required a density $> 10^7$, and perhaps much larger, for the time that the H$\alpha$ emission was prevalent, thus suggesting the potential for detectable X-ray emission. X-ray observations were conducted more than 4 years after explosion for both the SNe studied herein. No X-ray emission was detected at the position of either SN. It is likely that the density had dropped significantly by then. Using the method of \citet{kbn91} we calculate 3$\sigma$ upper limits to the X-ray emission from the SN to not exceed \textbf{2 $\times 10^{40}$} erg s$^{-1}$.  These are consistent with those obtained for SN 2018fhw, and lower than those for Type Ia-CSM SNe by at least an order of magnitude, if not more. Without additional information, it is difficult to place too many constraints on the properties of the medium around these SNe. We can say that the density decreased for SN 2016jae, although it is not so clear for SN 2018cqj.

X-ray observations of three low-luminosity Type Ia SNe with a H line in the optical spectrum have now resulted in no detection. Additionally, at least two Type Ia-CSM SNe, and numerous other Type Ia SNe, have been observed in X-rays without successfully being detected.  Another Type Ia SN, 2020uem, was found to show a H line in its optical spectrum, and estimated to be expanding in a high density medium \citep{unoetal23a, unoetal23b, sharmaetal23}. To date, only one Type Ia SN, SN 2012ca, has been detected in X-rays. 2020eyj, a Type Ia SN that showed He lines, but no H, was detected recently in the radio.  If all of these, including the Type Ia-CSMs, are proven to be bona-fide Type Ia SNe, they indicate that at least a fraction of Type Ia's may interact with a high density medium at some point early in their lifetime, contrary to the expectation that all Type Ia's are expanding in a low density medium. 

It is imperative that we explore and detect this population of Type Ia SNe with a H line in their spectrum, to get an estimate on the size of such a population compared to the total number of Type Ia SNe.  It is not clear whether all Type Ia's with a single degenerate progenitor would show an H$\alpha$ line. The detection of SN 2020eyj, with He but no H lines, suggests that there is considerable variety among the companion stars in single degenerate models, and thus in Type Ia SN progenitors. It is also possible that there may exist a number of Type Ia SNe where the ambient density is higher than average, but not high enough to generate a detectable H (or He) line when viewed with current telescopes. The only way to answer these questions is to find and study as many Ia's of this type as possible. This will help to constrain the fraction of Type Ia SNe that have single degenerate progenitors. When only Type Ia-CSM SNe were known to have a H line in the spectrum, it was sometimes argued that these were not bona-fide Type Ia's, or that they were extreme cases. These arguments cannot be applied to SN 2018fhw, SN 2018cqj and SN 2016jae, which are clearly low-luminosity Type Ia SNe. 

\citet{dwarkadas23a} discussed (in the Conclusions section) the timing of the X-ray observations of various SNe. This can be compared to when the H$\alpha$ line was observed, as is clearly seen in Figure~\ref{fig:snxray} and Figure~\ref{fig:snxray2}. In SN 2005gj and SN 2002ic, X-ray observations were made less than a year after explosion, and were almost co-incident with H$\alpha$ observations. However no X-ray emission was seen. The X-ray limits are higher for these 1a-CSM SNe, although still comparable to or lower than the H$\alpha$ luminosity. The expectation of high density would suggest the onset of detectable X-ray emission. Another factor in the detectability of the emission would be the ionization of the medium - a fully ionized medium would allow all the X-ray emission to escape, whereas a low ionization factor or neutral medium would result in absorption of the X-rays. One possibility is that when the X-ray observations were made, the column density was very high ($ > 1 \times 10^{23} $ cm$^{-2}$ ), the medium was mostly neutral further away from the SN, and any intrinsic X-ray emission was absorbed. In the case of the low luminosity SNe, it appears that the X-ray observations were made too late after the epoch of H$\alpha$ observations, and the ambient density had decreased to the point that there was no detectable X-ray emission. SN 2012ca was, perhaps serendipitously, observed at just the right time, when the density was still high, but not high enough that the emission was all absorbed, and the circumstellar medium was at least partially ionized. Arguably then, a case can be made that perhaps the optimal time of X-ray observations is when H$\alpha$ emission is still detectable but its luminosity has been declining for some time. In that case the density would still be high, and the X-ray emission detectable, but the column density will have reduced, and the medium is mostly ionized. Of course, it is difficult to decide quantitatively when this should be. 

Despite the non-detections, we assert that it is important to keep studying Type Ia SNe with an H$\alpha$ or He line at X-ray wavelengths. The detection of any Type Ia in X-rays is important. However it is the ones with a H (or He) line in their optical spectrum that are likely to have a high X-ray luminosity, allowing them to be detected using currently available X-ray instruments. Follow up at other wavelengths, especially radio, is also highly encouraged, as both of these would be good indicators of circumstellar interaction. The rewards of finding even a single confirmed Type Ia at these wavelengths one are significant, outweighing the observational risk of a non-detection.

\section*{Acknowledgements}
 Support for this work was provided by the National Aeronautics and Space Administration through Chandra Award Number GO3-24041X  issued by the Chandra X-ray Center, which is operated by the Smithsonian Astrophysical Observatory for and on behalf of the National Aeronautics Space Administration under contract NAS8-03060, and by National Science Foundation grant 1911061 awarded to the University of Chicago (PI: Vikram Dwarkadas). We thank the referee for a thorough reading of the paper, and for several useful comments and suggestions, which helped to considerably improve the manuscript. The scientific results reported in this article are based to a significant degree on observations made by the Chandra X-ray Observatory, including ObsIDs 26612, 26613, 27686, 27687, 27688. This research has made use of software provided by the Chandra X-ray Center (CXC) in the application packages CIAO and Sherpa. It has also made use of the Chandra PIMMS and Colden packages. 

\section*{Data Availability}

All data are included in the tables in the article.



\bibliographystyle{mnras}
\bibliography{halpha} 

\begin{thebibliography}{}
\makeatletter
\relax
\def\mn@urlcharsother{\let\do\@makeother \do\$\do\&\do\#\do\^\do\_\do\%\do\~}
\def\mn@doi{\begingroup\mn@urlcharsother \@ifnextchar [ {\mn@doi@}
  {\mn@doi@[]}}
\def\mn@doi@[#1]#2{\def\@tempa{#1}\ifx\@tempa\@empty \href
  {http://dx.doi.org/#2} {doi:#2}\else \href {http://dx.doi.org/#2} {#1}\fi
  \endgroup}
\def\mn@eprint#1#2{\mn@eprint@#1:#2::\@nil}
\def\mn@eprint@arXiv#1{\href {http://arxiv.org/abs/#1} {{\tt arXiv:#1}}}
\def\mn@eprint@dblp#1{\href {http://dblp.uni-trier.de/rec/bibtex/#1.xml}
  {dblp:#1}}
\def\mn@eprint@#1:#2:#3:#4\@nil{\def\@tempa {#1}\def\@tempb {#2}\def\@tempc
  {#3}\ifx \@tempc \@empty \let \@tempc \@tempb \let \@tempb \@tempa \fi \ifx
  \@tempb \@empty \def\@tempb {arXiv}\fi \@ifundefined
  {mn@eprint@\@tempb}{\@tempb:\@tempc}{\expandafter \expandafter \csname
  mn@eprint@\@tempb\endcsname \expandafter{\@tempc}}}

\bibitem[\protect\citeauthoryear{{Bietenholz}, {Bartel}, {Argo}, {Dua}, {Ryder}
   \& {Soderberg}}{{Bietenholz} et~al.}{2021}]{bietenholzetal21}
{Bietenholz} M.~F.,  {Bartel} N.,  {Argo} M.,  {Dua} R.,  {Ryder} S.,
  {Soderberg} A.,  2021, \mn@doi [\apj] {10.3847/1538-4357/abccd9}, \href
  {https://ui.adsabs.harvard.edu/abs/2021ApJ...908...75B} {908, 75}

\bibitem[\protect\citeauthoryear{{Bloom} et~al.,}{{Bloom}
  et~al.}{2012}]{bloometal12}
{Bloom} J.~S.,  et~al., 2012, \mn@doi [\apjl] {10.1088/2041-8205/744/2/L17},
  \href {https://ui.adsabs.harvard.edu/abs/2012ApJ...744L..17B} {744, L17}

\bibitem[\protect\citeauthoryear{{Bochenek}, {Dwarkadas}, {Silverman}, {Fox},
  {Chevalier}, {Smith}  \& {Filippenko}}{{Bochenek}
  et~al.}{2018}]{bocheneketal18}
{Bochenek} C.~D.,  {Dwarkadas} V.~V.,  {Silverman} J.~M.,  {Fox} O.~D.,
  {Chevalier} R.~A.,  {Smith} N.,   {Filippenko} A.~V.,  2018, \mn@doi [\mnras]
  {10.1093/mnras/stx2029}, \href
  {https://ui.adsabs.harvard.edu/abs/2018MNRAS.473..336B} {473, 336}

\bibitem[\protect\citeauthoryear{{Brown} et~al.,}{{Brown}
  et~al.}{2012}]{brownetal12}
{Brown} P.~J.,  et~al., 2012, \mn@doi [\apj] {10.1088/0004-637X/753/1/22},
  \href {https://ui.adsabs.harvard.edu/abs/2012ApJ...753...22B} {753, 22}

\bibitem[\protect\citeauthoryear{{Chambers} et~al.,}{{Chambers}
  et~al.}{2016}]{chambersetal16}
{Chambers} K.~C.,  et~al., 2016, \mn@doi [arXiv e-prints]
  {10.48550/arXiv.1612.05560}, \href
  {https://ui.adsabs.harvard.edu/abs/2016arXiv161205560C} {p. arXiv:1612.05560}

\bibitem[\protect\citeauthoryear{{Chevalier} \& {Fransson}}{{Chevalier} \&
  {Fransson}}{1994}]{cf94}
{Chevalier} R.~A.,  {Fransson} C.,  1994, \mn@doi [\apj] {10.1086/173557},
  \href {https://ui.adsabs.harvard.edu/abs/1994ApJ...420..268C} {420, 268}

\bibitem[\protect\citeauthoryear{{Chevalier} \& {Raymond}}{{Chevalier} \&
  {Raymond}}{1978}]{cr78}
{Chevalier} R.~A.,  {Raymond} J.~C.,  1978, \mn@doi [\apjl] {10.1086/182785},
  \href {https://ui.adsabs.harvard.edu/abs/1978ApJ...225L..27C} {225, L27}

\bibitem[\protect\citeauthoryear{{Chugai}}{{Chugai}}{1993}]{chugai93}
{Chugai} N.~N.,  1993, \mn@doi [\apjl] {10.1086/187006}, \href
  {https://ui.adsabs.harvard.edu/abs/1993ApJ...414L.101C} {414, L101}

\bibitem[\protect\citeauthoryear{{Chugai} \& {Danziger}}{{Chugai} \&
  {Danziger}}{1994}]{cd94}
{Chugai} N.~N.,  {Danziger} I.~J.,  1994, \mn@doi [\mnras]
  {10.1093/mnras/268.1.173}, \href
  {https://ui.adsabs.harvard.edu/abs/1994MNRAS.268..173C} {268, 173}

\bibitem[\protect\citeauthoryear{{Deng} et~al.,}{{Deng}
  et~al.}{2004}]{dengetal04}
{Deng} J.,  et~al., 2004, \mn@doi [\apjl] {10.1086/420698}, \href
  {https://ui.adsabs.harvard.edu/abs/2004ApJ...605L..37D} {605, L37}

\bibitem[\protect\citeauthoryear{{Dickey} \& {Lockman}}{{Dickey} \&
  {Lockman}}{1990}]{dl90}
{Dickey} J.~M.,  {Lockman} F.~J.,  1990, \mn@doi [\araa]
  {10.1146/annurev.aa.28.090190.001243}, \href
  {https://ui.adsabs.harvard.edu/abs/1990ARA&A..28..215D} {28, 215}

\bibitem[\protect\citeauthoryear{{Drake} \& {Ulrich}}{{Drake} \&
  {Ulrich}}{1980}]{du80}
{Drake} S.~A.,  {Ulrich} R.~K.,  1980, \mn@doi [\apjs] {10.1086/190654}, \href
  {https://ui.adsabs.harvard.edu/abs/1980ApJS...42..351D} {42, 351}

\bibitem[\protect\citeauthoryear{{Dwarkadas}}{{Dwarkadas}}{2014}]{vvd14}
{Dwarkadas} V.~V.,  2014, \mn@doi [\mnras] {10.1093/mnras/stu347}, \href
  {https://ui.adsabs.harvard.edu/abs/2014MNRAS.440.1917D} {440, 1917}

\bibitem[\protect\citeauthoryear{{Dwarkadas}}{{Dwarkadas}}{2023a}]{dwarkadas23b}
{Dwarkadas} V.~V.,  2023a, \mn@doi [Research Notes of the American Astronomical
  Society] {10.3847/2515-5172/acdf53}, \href
  {https://ui.adsabs.harvard.edu/abs/2023RNAAS...7..129D} {7, 129}

\bibitem[\protect\citeauthoryear{{Dwarkadas}}{{Dwarkadas}}{2023b}]{dwarkadas23a}
{Dwarkadas} V.~V.,  2023b, \mn@doi [\mnras] {10.1093/mnras/stac3384}, \href
  {https://ui.adsabs.harvard.edu/abs/2023MNRAS.520.1362D} {520, 1362}

\bibitem[\protect\citeauthoryear{{Dwarkadas} \& {Gruszko}}{{Dwarkadas} \&
  {Gruszko}}{2012}]{dg12}
{Dwarkadas} V.~V.,  {Gruszko} J.,  2012, \mn@doi [\mnras]
  {10.1111/j.1365-2966.2011.19808.x}, \href
  {https://ui.adsabs.harvard.edu/abs/2012MNRAS.419.1515D} {419, 1515}

\bibitem[\protect\citeauthoryear{{Dwarkadas}, {Romero-Ca{\~n}izales}, {Reddy}
  \& {Bauer}}{{Dwarkadas} et~al.}{2016}]{drrb16}
{Dwarkadas} V.~V.,  {Romero-Ca{\~n}izales} C.,  {Reddy} R.,   {Bauer} F.~E.,
  2016, \mn@doi [\mnras] {10.1093/mnras/stw1717}, \href
  {https://ui.adsabs.harvard.edu/abs/2016MNRAS.462.1101D} {462, 1101}

\bibitem[\protect\citeauthoryear{{Elias-Rosa} et~al.,}{{Elias-Rosa}
  et~al.}{2021}]{eliasrosaetal21}
{Elias-Rosa} N.,  et~al., 2021, \mn@doi [\aap] {10.1051/0004-6361/202141218},
  \href {https://ui.adsabs.harvard.edu/abs/2021A&A...652A.115E} {652, A115}

\bibitem[\protect\citeauthoryear{{Fox} \& {Filippenko}}{{Fox} \&
  {Filippenko}}{2013}]{ff13}
{Fox} O.~D.,  {Filippenko} A.~V.,  2013, \mn@doi [\apjl]
  {10.1088/2041-8205/772/1/L6}, \href
  {https://ui.adsabs.harvard.edu/abs/2013ApJ...772L...6F} {772, L6}

\bibitem[\protect\citeauthoryear{{Fox} et~al.,}{{Fox} et~al.}{2015}]{foxetal15}
{Fox} O.~D.,  et~al., 2015, \mn@doi [\mnras] {10.1093/mnras/stu2435}, \href
  {https://ui.adsabs.harvard.edu/abs/2015MNRAS.447..772F} {447, 772}

\bibitem[\protect\citeauthoryear{{Graham} et~al.,}{{Graham}
  et~al.}{2019}]{grahametal19}
{Graham} M.~L.,  et~al., 2019, \mn@doi [\apj] {10.3847/1538-4357/aaf41e}, \href
  {https://ui.adsabs.harvard.edu/abs/2019ApJ...871...62G} {871, 62}

\bibitem[\protect\citeauthoryear{{Gress} et~al.,}{{Gress}
  et~al.}{2016}]{gressetal16}
{Gress} O.,  et~al., 2016, The Astronomer's Telegram, \href
  {https://ui.adsabs.harvard.edu/abs/2016ATel.9902....1G} {9902, 1}

\bibitem[\protect\citeauthoryear{{Hamuy} et~al.,}{{Hamuy}
  et~al.}{2003}]{hamuyetal03}
{Hamuy} M.,  et~al., 2003, \mn@doi [\nat] {10.1038/nature01854}, \href
  {https://ui.adsabs.harvard.edu/abs/2003Natur.424..651H} {424, 651}

\bibitem[\protect\citeauthoryear{{Hodgkin}, {Wyrzykowski}, {Blagorodnova}  \&
  {Koposov}}{{Hodgkin} et~al.}{2013}]{hodgkinetal13}
{Hodgkin} S.~T.,  {Wyrzykowski} L.,  {Blagorodnova} N.,   {Koposov} S.,  2013,
  \mn@doi [Philosophical Transactions of the Royal Society of London Series A]
  {10.1098/rsta.2012.0239}, \href
  {https://ui.adsabs.harvard.edu/abs/2013RSPTA.37120239H} {371, 20120239}

\bibitem[\protect\citeauthoryear{{Horesh} et~al.,}{{Horesh}
  et~al.}{2012}]{horeshetal12}
{Horesh} A.,  et~al., 2012, \mn@doi [\apj] {10.1088/0004-637X/746/1/21}, \href
  {https://ui.adsabs.harvard.edu/abs/2012ApJ...746...21H} {746, 21}

\bibitem[\protect\citeauthoryear{{Hughes}, {Chugai}, {Chevalier}, {Lundqvist}
  \& {Schlegel}}{{Hughes} et~al.}{2007}]{hughesetal07}
{Hughes} J.~P.,  {Chugai} N.,  {Chevalier} R.,  {Lundqvist} P.,   {Schlegel}
  E.,  2007, \mn@doi [\apj] {10.1086/522113}, \href
  {https://ui.adsabs.harvard.edu/abs/2007ApJ...670.1260H} {670, 1260}

\bibitem[\protect\citeauthoryear{{Inserra} et~al.,}{{Inserra}
  et~al.}{2014}]{inserraetal14}
{Inserra} C.,  et~al., 2014, \mn@doi [\mnras] {10.1093/mnrasl/slt138}, \href
  {https://ui.adsabs.harvard.edu/abs/2014MNRAS.437L..51I} {437, L51}

\bibitem[\protect\citeauthoryear{{Kollmeier} et~al.,}{{Kollmeier}
  et~al.}{2019}]{kollmeieretal19}
{Kollmeier} J.~A.,  et~al., 2019, \mn@doi [\mnras] {10.1093/mnras/stz953},
  \href {https://ui.adsabs.harvard.edu/abs/2019MNRAS.486.3041K} {486, 3041}

\bibitem[\protect\citeauthoryear{{Kool} et~al.,}{{Kool}
  et~al.}{2023}]{kooletal23}
{Kool} E.~C.,  et~al., 2023, \mn@doi [\nat] {10.1038/s41586-023-05916-w}, \href
  {https://ui.adsabs.harvard.edu/abs/2023Natur.617..477K} {617, 477}

\bibitem[\protect\citeauthoryear{{Kraft}, {Burrows}  \& {Nousek}}{{Kraft}
  et~al.}{1991}]{kbn91}
{Kraft} R.~P.,  {Burrows} D.~N.,   {Nousek} J.~A.,  1991, \mn@doi [\apj]
  {10.1086/170124}, \href
  {https://ui.adsabs.harvard.edu/abs/1991ApJ...374..344K} {374, 344}

\bibitem[\protect\citeauthoryear{{Liu}, {R{\"o}pke}  \& {Han}}{{Liu}
  et~al.}{2023}]{lrh23}
{Liu} Z.-W.,  {R{\"o}pke} F.~K.,   {Han} Z.,  2023, \mn@doi [Research in
  Astronomy and Astrophysics] {10.1088/1674-4527/acd89e}, \href
  {https://ui.adsabs.harvard.edu/abs/2023RAA....23h2001L} {23, 082001}

\bibitem[\protect\citeauthoryear{{Livio} \& {Mazzali}}{{Livio} \&
  {Mazzali}}{2018}]{lm18}
{Livio} M.,  {Mazzali} P.,  2018, \mn@doi [\physrep]
  {10.1016/j.physrep.2018.02.002}, \href
  {https://ui.adsabs.harvard.edu/abs/2018PhR...736....1L} {736, 1}

\bibitem[\protect\citeauthoryear{{Lundqvist} et~al.,}{{Lundqvist}
  et~al.}{2013}]{lundqvistetal13}
{Lundqvist} P.,  et~al., 2013, \mn@doi [\mnras] {10.1093/mnras/stt1303}, \href
  {https://ui.adsabs.harvard.edu/abs/2013MNRAS.435..329L} {435, 329}

\bibitem[\protect\citeauthoryear{{Magnier} et~al.,}{{Magnier}
  et~al.}{2020}]{magnieretal20}
{Magnier} E.~A.,  et~al., 2020, \mn@doi [\apjs] {10.3847/1538-4365/abb82a},
  \href {https://ui.adsabs.harvard.edu/abs/2020ApJS..251....6M} {251, 6}

\bibitem[\protect\citeauthoryear{{Maoz}, {Mannucci}  \& {Nelemans}}{{Maoz}
  et~al.}{2014}]{maoz14}
{Maoz} D.,  {Mannucci} F.,   {Nelemans} G.,  2014, \mn@doi [\araa]
  {10.1146/annurev-astro-082812-141031}, \href
  {https://ui.adsabs.harvard.edu/abs/2014ARA&A..52..107M} {52, 107}

\bibitem[\protect\citeauthoryear{{Margutti}, {Parrent}, {Kamble}, {Soderberg},
  {Foley}, {Milisavljevic}, {Drout}  \& {Kirshner}}{{Margutti}
  et~al.}{2014}]{marguttietal14}
{Margutti} R.,  {Parrent} J.,  {Kamble} A.,  {Soderberg} A.~M.,  {Foley} R.~J.,
   {Milisavljevic} D.,  {Drout} M.~R.,   {Kirshner} R.,  2014, \mn@doi [\apj]
  {10.1088/0004-637X/790/1/52}, \href
  {https://ui.adsabs.harvard.edu/abs/2014ApJ...790...52M} {790, 52}

\bibitem[\protect\citeauthoryear{{Nugent} et~al.,}{{Nugent}
  et~al.}{2011}]{nugentetal11}
{Nugent} P.~E.,  et~al., 2011, \mn@doi [\nat] {10.1038/nature10644}, \href
  {https://ui.adsabs.harvard.edu/abs/2011Natur.480..344N} {480, 344}

\bibitem[\protect\citeauthoryear{{Patat} et~al.,}{{Patat}
  et~al.}{2007}]{patatetal07}
{Patat} F.,  et~al., 2007, \mn@doi [Science] {10.1126/science.1143005}, \href
  {https://ui.adsabs.harvard.edu/abs/2007Sci...317..924P} {317, 924}

\bibitem[\protect\citeauthoryear{{Prieto} et~al.,}{{Prieto}
  et~al.}{2020}]{prietoetal20}
{Prieto} J.~L.,  et~al., 2020, \mn@doi [\apj] {10.3847/1538-4357/ab6323}, \href
  {https://ui.adsabs.harvard.edu/abs/2020ApJ...889..100P} {889, 100}

\bibitem[\protect\citeauthoryear{{R{\"o}pke} \& {Sim}}{{R{\"o}pke} \&
  {Sim}}{2018}]{rs18}
{R{\"o}pke} F.~K.,  {Sim} S.~A.,  2018, \mn@doi [\ssr]
  {10.1007/s11214-018-0503-8}, \href
  {https://ui.adsabs.harvard.edu/abs/2018SSRv..214...72R} {214, 72}

\bibitem[\protect\citeauthoryear{{Ross} \& {Dwarkadas}}{{Ross} \&
  {Dwarkadas}}{2017}]{rd17}
{Ross} M.,  {Dwarkadas} V.~V.,  2017, \mn@doi [\aj] {10.3847/1538-3881/aa6d50},
  \href {https://ui.adsabs.harvard.edu/abs/2017AJ....153..246R} {153, 246}

\bibitem[\protect\citeauthoryear{{Ruiter}}{{Ruiter}}{2020}]{ruiter20}
{Ruiter} A.~J.,  2020, \mn@doi [IAU Symposium] {10.1017/S1743921320000587},
  \href {https://ui.adsabs.harvard.edu/abs/2020IAUS..357....1R} {357, 1}

\bibitem[\protect\citeauthoryear{{Ruiz-Lapuente}}{{Ruiz-Lapuente}}{2014}]{pilar14}
{Ruiz-Lapuente} P.,  2014, \mn@doi [\nar] {10.1016/j.newar.2014.08.002}, \href
  {https://ui.adsabs.harvard.edu/abs/2014NewAR..62...15R} {62, 15}

\bibitem[\protect\citeauthoryear{{Russell} \& {Immler}}{{Russell} \&
  {Immler}}{2012}]{ri12}
{Russell} B.~R.,  {Immler} S.,  2012, \mn@doi [\apjl]
  {10.1088/2041-8205/748/2/L29}, \href
  {https://ui.adsabs.harvard.edu/abs/2012ApJ...748L..29R} {748, L29}

\bibitem[\protect\citeauthoryear{{Sand} et~al.,}{{Sand}
  et~al.}{2021}]{sandetal21}
{Sand} D.~J.,  et~al., 2021, \mn@doi [\apj] {10.3847/1538-4357/ac20da}, \href
  {https://ui.adsabs.harvard.edu/abs/2021ApJ...922...21S} {922, 21}

\bibitem[\protect\citeauthoryear{{Scalzo} et~al.,}{{Scalzo}
  et~al.}{2010}]{scalzoetal10}
{Scalzo} R.~A.,  et~al., 2010, \mn@doi [\apj] {10.1088/0004-637X/713/2/1073},
  \href {https://ui.adsabs.harvard.edu/abs/2010ApJ...713.1073S} {713, 1073}

\bibitem[\protect\citeauthoryear{{Schlegel} \& {Petre}}{{Schlegel} \&
  {Petre}}{1993}]{sp93}
{Schlegel} E.~M.,  {Petre} R.,  1993, \mn@doi [\apjl] {10.1086/186932}, \href
  {https://ui.adsabs.harvard.edu/abs/1993ApJ...412L..29S} {412, L29}

\bibitem[\protect\citeauthoryear{{Sharma} et~al.,}{{Sharma}
  et~al.}{2023}]{sharmaetal23}
{Sharma} Y.,  et~al., 2023, \mn@doi [\apj] {10.3847/1538-4357/acbc16}, \href
  {https://ui.adsabs.harvard.edu/abs/2023ApJ...948...52S} {948, 52}

\bibitem[\protect\citeauthoryear{{Silverman}, {Ganeshalingam}, {Li},
  {Filippenko}, {Miller}  \& {Poznanski}}{{Silverman}
  et~al.}{2011}]{silvermanetal11}
{Silverman} J.~M.,  {Ganeshalingam} M.,  {Li} W.,  {Filippenko} A.~V.,
  {Miller} A.~A.,   {Poznanski} D.,  2011, \mn@doi [\mnras]
  {10.1111/j.1365-2966.2010.17474.x}, \href
  {https://ui.adsabs.harvard.edu/abs/2011MNRAS.410..585S} {410, 585}

\bibitem[\protect\citeauthoryear{{Silverman} et~al.,}{{Silverman}
  et~al.}{2013}]{silvermanetal13}
{Silverman} J.~M.,  et~al., 2013, \mn@doi [\apjs] {10.1088/0067-0049/207/1/3},
  \href {https://ui.adsabs.harvard.edu/abs/2013ApJS..207....3S} {207, 3}

\bibitem[\protect\citeauthoryear{{Smith} et~al.,}{{Smith}
  et~al.}{2016a}]{smithetal16b}
{Smith} K.~W.,  et~al., 2016a, The Astronomer's Telegram, \href
  {https://ui.adsabs.harvard.edu/abs/2016ATel.9908....1S} {9908, 1}

\bibitem[\protect\citeauthoryear{{Smith}, {Cikota}, {Magee}, {Kankare}  \&
  {Yaron}}{{Smith} et~al.}{2016b}]{smithetal16a}
{Smith} K.,  {Cikota} A.,  {Magee} M.,  {Kankare} E.,   {Yaron} O.,  2016b,
  Transient Name Server Classification Report, \href
  {https://ui.adsabs.harvard.edu/abs/2016TNSCR1095....1S} {2016-1095, 1}

\bibitem[\protect\citeauthoryear{{Soker}}{{Soker}}{2019}]{soker19}
{Soker} N.,  2019, \mn@doi [\nar] {10.1016/j.newar.2020.101535}, \href
  {https://ui.adsabs.harvard.edu/abs/2019NewAR..8701535S} {87, 101535}

\bibitem[\protect\citeauthoryear{{Soker}, {Garc{\'\i}a-Berro}  \&
  {Althaus}}{{Soker} et~al.}{2014}]{sokeretal14}
{Soker} N.,  {Garc{\'\i}a-Berro} E.,   {Althaus} L.~G.,  2014, \mn@doi [\mnras]
  {10.1093/mnrasl/slt142}, \href
  {https://ui.adsabs.harvard.edu/abs/2014MNRAS.437L..66S} {437, L66}

\bibitem[\protect\citeauthoryear{{Tanikawa}, {Nomoto}, {Nakasato}  \&
  {Maeda}}{{Tanikawa} et~al.}{2019}]{tn19}
{Tanikawa} A.,  {Nomoto} K.,  {Nakasato} N.,   {Maeda} K.,  2019, \mn@doi
  [\apj] {10.3847/1538-4357/ab46b6}, \href
  {https://ui.adsabs.harvard.edu/abs/2019ApJ...885..103T} {885, 103}

\bibitem[\protect\citeauthoryear{{Taubenberger}}{{Taubenberger}}{2017}]{taubenberger17}
{Taubenberger} S.,  2017, in {Alsabti} A.~W.,  {Murdin} P.,  eds, , Handbook of
  Supernovae.
Springer International Publishing, p.~317,
  \mn@doi{10.1007/978-3-319-21846-5\_37}

\bibitem[\protect\citeauthoryear{{Tonry} et~al.,}{{Tonry}
  et~al.}{2018}]{tonryetal18}
{Tonry} J.~L.,  et~al., 2018, \mn@doi [\pasp] {10.1088/1538-3873/aabadf}, \href
  {https://ui.adsabs.harvard.edu/abs/2018PASP..130f4505T} {130, 064505}

\bibitem[\protect\citeauthoryear{{Tsebrenko} \& {Soker}}{{Tsebrenko} \&
  {Soker}}{2015}]{ts15}
{Tsebrenko} D.,  {Soker} N.,  2015, \mn@doi [\mnras] {10.1093/mnras/stu2567},
  \href {https://ui.adsabs.harvard.edu/abs/2015MNRAS.447.2568T} {447, 2568}

\bibitem[\protect\citeauthoryear{{Uno} et~al.,}{{Uno}
  et~al.}{2023a}]{unoetal23a}
{Uno} K.,  et~al., 2023a, \mn@doi [\apj] {10.3847/1538-4357/acb5ec}, \href
  {https://ui.adsabs.harvard.edu/abs/2023ApJ...944..203U} {944, 203}

\bibitem[\protect\citeauthoryear{{Uno} et~al.,}{{Uno}
  et~al.}{2023b}]{unoetal23b}
{Uno} K.,  et~al., 2023b, \mn@doi [\apj] {10.3847/1538-4357/acb5eb}, \href
  {https://ui.adsabs.harvard.edu/abs/2023ApJ...944..204U} {944, 204}

\bibitem[\protect\citeauthoryear{{Vallely} et~al.,}{{Vallely}
  et~al.}{2019}]{vallelyetal19}
{Vallely} P.~J.,  et~al., 2019, \mn@doi [\mnras] {10.1093/mnras/stz1445}, \href
  {https://ui.adsabs.harvard.edu/abs/2019MNRAS.487.2372V} {487, 2372}

\bibitem[\protect\citeauthoryear{{Wang} et~al.,}{{Wang}
  et~al.}{2009}]{wangetal09}
{Wang} X.,  et~al., 2009, \mn@doi [\apjl] {10.1088/0004-637X/699/2/L139}, \href
  {https://ui.adsabs.harvard.edu/abs/2009ApJ...699L.139W} {699, L139}

\bibitem[\protect\citeauthoryear{{Weaver}, {McCray}, {Castor}, {Shapiro}  \&
  {Moore}}{{Weaver} et~al.}{1977}]{weaveretal77}
{Weaver} R.,  {McCray} R.,  {Castor} J.,  {Shapiro} P.,   {Moore} R.,  1977,
  \mn@doi [\apj] {10.1086/155692}, \href
  {https://ui.adsabs.harvard.edu/abs/1977ApJ...218..377W} {218, 377}

\makeatother
\end{thebibliography}




\bsp	
\label{lastpage}
\end{document}